# On-chip thermo-optic tuning of suspended microresonators

**BRIAN S. LEE[1], MIAN ZHANG[2], FELIPPE A. S. BARBOSA[1], STEVEN A. MILLER[1,3], ASEEMA MOHANTY[1,3], RAPHAEL ST-GELAIS[1,3], AND MICHAL LIPSON[1,4]**

[1]*Department of Electrical Engineering, Columbia University, New York, NY 10027, USA*
[2]*School of Engineering and Applied Sciences, Harvard University, Cambridge, Massachusetts 02138, USA*
[3]*School of Electrical and Computer Engineering, Cornell University, Ithaca, New York 14853, USA*
[4]*ml3745@columbia.edu*

**Abstract:** Suspended optical microresonators are promising devices for on-chip photonic applications such as radio-frequency oscillators, optical frequency combs, and sensors. Scaling up these devices demand the capability to tune the optical resonances in an integrated manner. Here, we design and experimentally demonstrate integrated on-chip thermo-optic tuning of suspended microresonators by utilizing suspended wire bridges and microheaters. We demonstrate the ability to tune the resonance of a suspended microresonator in silicon nitride platform by 9.7 GHz using 5.3 mW of heater power. The loaded optical quality factor ($Q_L \sim 92{,}000$) stays constant throughout the detuning. We demonstrate the efficacy of our approach by completely turning on and off the optical coupling between two evanescently coupled suspended microresonators.

## 1. Introduction

Microresonators are the workhorse for diverse on-chip photonics applications from biochemical sensing [1, 2], non-linear optics [3, 4], optomechanics [5–7], to on-chip lasers [8–10]. These applications increasingly utilize air-cladded and suspended microresonator geometry to leverage the strong light confinement from the high index contrast, the direct interaction between the sensing targets and the optical field, and the transduction between optics and mechanics via radiation pressure.

Many suspended microresonator-based applications demand the ability to actively tune the optical resonant frequencies to compensate for the unavoidable fabrication variations, but the lack of an efficient integrated tuning scheme for these resonators has significantly limited their use. The difficulty of tuning air-cladded resonators in an integrated manner arises from the lack of optical isolation between a tuning mechanism and the optical mode. To illustrate this, consider tuning a ring resonator through thermo-optic effect using heaters [11, 12]. Typically, a microheater is positioned on a cladding layer above the optical devices to isolate the metal from the optical field underneath. For suspended microresonators, however, the heaters cannot be positioned directly above the optical devices because they are air-cladded and the metal will absorb light.

Previous efforts to tune suspended microresonators focus on isolating the tuning mechanism from the optical field without sacrificing efficiency, but they are either fundamentally restricted to tuning a single resonator or their tuning mechanisms are not fully integrated. For example, the resonance of a suspended microresonator was tuned by photo-electrochemical etching [13], but tuning through fabrication does not allow reconfiguration. Several groups have demonstrated thermo-optic tuning with Peltier element stage heating [14, 15], but this approach is inefficient and non-local, where multiple resonators cannot be tuned independently. There has also been resonance tuning using carrier-injection [16], but this effect is inherently associated with optical loss and is limited to active materials such as silicon, germanium, or other compound semiconductors. Another approach is using an out-of-plane laser tuning [5, 17], but this is inherently non-scalable as this approach would require one laser per resonator. Other methods include electrically tuning suspended microresonators, such as capacitive strain-induced tuning [18] and resistive tuning [19]. Currently, however, the electrical probes directly touch the resonators, making them non-scalable and non-integrated. To control the resonance dynamics on-demand and in an integrated manner, the tuning mechanisms should be controlled with integrated wires and contact pads isolated from the device.

Here, we demonstrate integrated thermo-optic tuning of suspended microresonators using suspended wire bridges and microheaters. Fig. 1 is a schematic of our approach. It consists of a resistive microheater placed at the center of the resonator as shown in Fig. 1(a). This allows efficient heat transfer through direct thermal contact while spatially isolating it from the resonator's optical whispering gallery mode at the perimeter. The heater is then connected to the electrical contacts isolated several tens of microns away from the resonators via suspended metal wire bridges. These suspended wires close the electrical connection between the microheater and the contacts without requiring a cladding material. As shown in Fig. 1(b), there is a small gap between the suspended wire bridges and the resonator edge to avoid metal absorbing light. Fig.1(c) describes how our approach can be scaled up for two evanescently coupled suspended resonators to control the coupling dynamics of multiple resonators in an integrated manner.

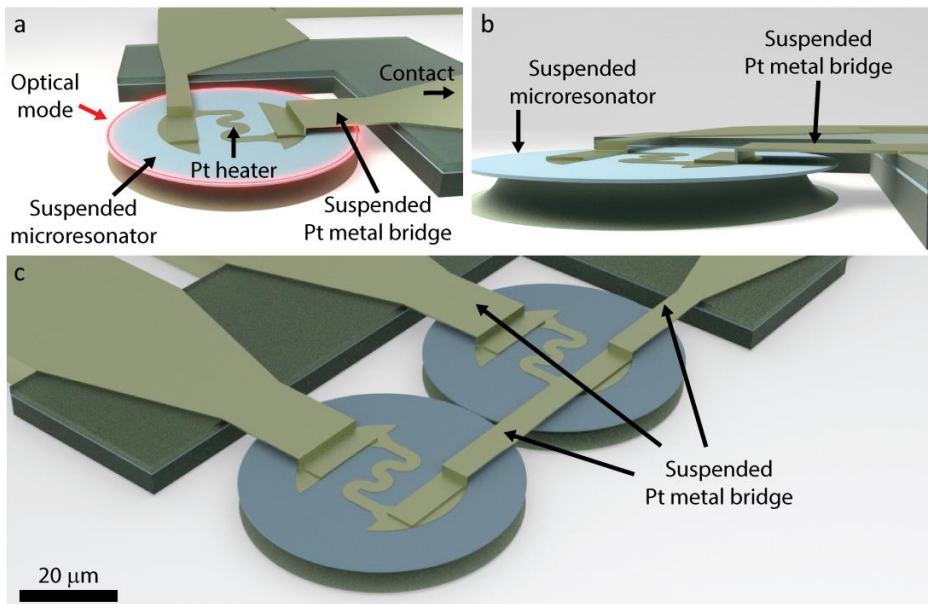

Fig. 1. Schematics of our integrated tuning of suspended microresonators. The suspended wire bridges are used to control the microheaters fabricated on the resonators without a cladding material. (a) We deposit a platinum microheater at the center of the resonator to achieve direct thermal contact for efficient heat transfer while isolating it from the optical mode at the resonator perimeter. The heater is connected to the contacts located tens of microns away from the resonator via suspended platinum wire bridges. The red perimeter depicts the optical whispering gallery mode of the resonator. (b) The side view of the device shows how the suspension of these wire bridges can electrically connect the heater to the contacts without a cladding layer and simultaneously avoid optical absorption. (c) An example of how this tuning approach can be scaled up for an array of two evanescently coupled suspended microresonators. Our approach allows independent tuning of each resonator to turn the optical coupling on and off on-demand. A scale bar is shown for reference.

## 2. Device fabrication and design

We fabricate our tunable suspended microresonators by depositing, annealing, and patterning optical structures in silicon nitride ($Si_3N_4$) and silicon oxide ($SiO_2$), followed by the depositing, patterning, and lifting-off metal structures for heaters and suspended wire bridges. Fig. 2 outlines the fabrication steps. We first deposit 240/200/220 nm of $Si_3N_4$/$SiO_2$/$Si_3N_4$ film stack on buried oxide silicon wafer (3 μm thermal $SiO_2$). The stoichiometric $Si_3N_4$ films are deposited by LPCVD with $SiO_2$ by PECVD and annealed for more than one hour at 1100 °C with $N_2$. We then pattern the resonator and support structures for the fiber taper and contacts with e-beam lithography as shown in Fig. 2(a). We etch $Si_3N_4$ with $CHF_3/O_2$ RIE plasma etch. We strip off the e-beam resist and clean the chip in piranha acid (3:1 ratio of sulfuric acid to hydrogen peroxide). Once the optical resonator is fabricated, we fabricate microheaters on top of the resonators. We deposit 10 nm chrome (Cr) adhesion layer followed by 100 nm of platinum (Pt) using sputter-deposition over the heater and contact resist patterns defined from contact photolithography. We lift-off excess metal and photoresist. We choose Pt and Cr adhesion layer as the main metal for the heater, and later for the suspended wires, because Pt and Cr are resistant to Buffered Oxide Etch (BOE, buffered hydrofluoric acid), which we later use to release the resonators. In addition, Pt is an appropriate metal for microheaters as it exhibits stable operating temperatures up to 450 °C [20]. After heater deposition, we deposit 1.2 μm thick PECVD $SiO_2$, which is later served as the sacrificial layer for the suspended metal bridges. The thickness of this sacrificial PECVD $SiO_2$ layer defines the separation gap between the suspended metal and the resonator (see Fig. 2(c), 2(e), and 2(f)). To ensure that the metal bridges do not interact with the optical fields of the resonator, the bridges are suspended more than one micron away from the device. Fig. 3 shows the simulated optical absorption loss (and reduction of optical quality factor, Q) caused by the suspended metal above the resonator optical mode with respect to the gap between them. This gap can be adjusted by varying the thickness of the PECVD $SiO_2$ layer. For our suspended microresonators with measured loaded Q in the order of $10^5$ and using Pt as the suspended metal, a gap greater than 1.2 μm provides sufficient optical isolation (simulated absorption losses of about 0.2 dB/cm, see Fig. 3 Left). We use PECVD $SiO_2$ as the sacrificial layer for the metal bridges to release the them simultaneously with the resonator in one BOE wet etching step. We pattern and etch the vias on the PECVD $SiO_2$ sacrificial layer with contact photolithography and RIE plasma $CHF_3/O_2$ chemistry, respectively, to expose the heater and contacts (Fig. 2(d)). After etching the vias, we pattern the bridges with contact lithography, and sputter-deposit and lift-off 10 nm Cr adhesion layer and 350 nm Pt. In this step, we engineer the stress of the deposited Pt films to minimize the optical absorption and maximize structural stability of the suspended wire bridges. We use the argon pressure (about 12 mTorr) during the sputter-deposition that yields Pt film with a slightly tensile stress of about 100 MPa. The stress of the deposited Pt film can be controlled by varying the argon pressure during sputtering deposition [22]. We fabricate the metal bridges with tensile stress to ensure that they do not buckle down after they are released with the resonator, as buckling can collapse the bridges or induce optical absorption. Also, as the tensile stress tends to bend the bridges upwards away from the device, it provides firm mechanical support for the suspended bridges. In order to find the optimal argon pressure yielding the appropriate tensile stress, we sputter-deposit 350 nm of Pt at different argon pressures on bare silicon wafers and measure the resulting film's stress by comparing the wafer bowing before and after the deposition. The tensile stress of about 100 MPa is sufficient to ensure structural stability and adequate adhesion of the metal after deposition. After lifting-off the final metal layer, we release the resonator perimeter and bridge wires by wet etching the sacrificial PECVD and thermal $SiO_2$ layer in BOE (6:1) for 35 minutes. We use a critical point dryer to minimize evaporation damages. Fig. 4 shows the fabricated suspended microresonators with the integrated tuning circuit taken with the scanning electron microscope (SEM).

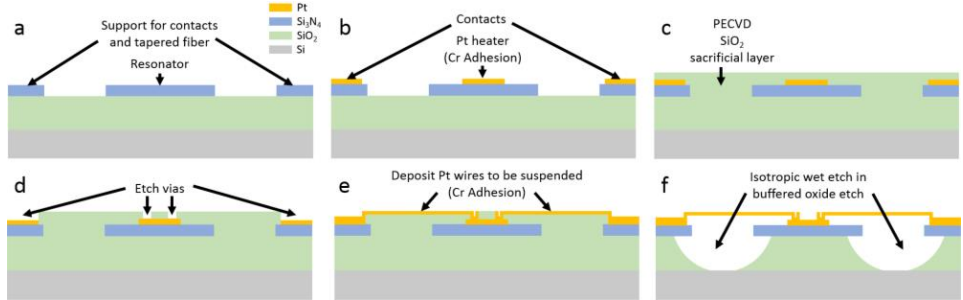

Fig. 2. The outline for fabricating tunable suspended microresonators with suspended wire bridges and microheaters. (a) The resonator and support structures for the electrical contacts and the fiber taper is defined using e-beam lithography. (b) Pt heater and contacts are sputter-deposited using contact lithography and lift-off. (c) The device is cladded with PECVD $SiO_2$, later to be served as a sacrificial layer for the suspended Pt wire bridges. (d) Vias are etched on the PECVD $SiO_2$ sacrificial layer to expose the microheater and the contacts for electrical connection. (e) Pt bridge wires are sputter-deposited to connect the heater and the contacts through the vias. The argon pressure during the deposition of the Pt film is chosen to ensure that the deposited metal film exhibits a tensile stress of about 100 MPa. (f) The resonator and the suspended wires are released simultaneously by wet etching in BOE. Critical point dryer is used to minimize evaporation damages.

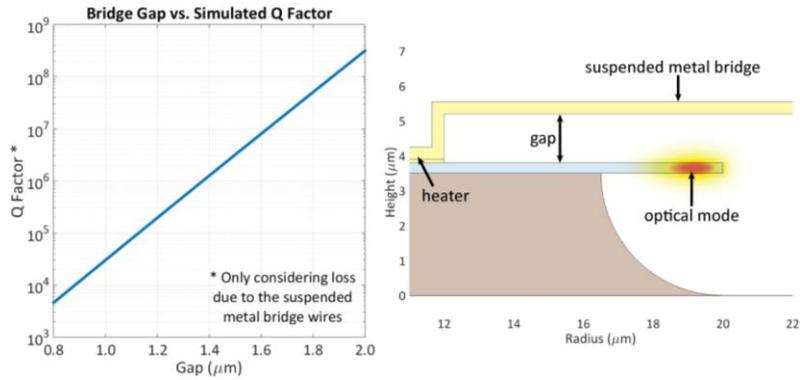

Fig. 3. (Left) Simulated Q factor limited by absorption loss from the metal suspended above the resonator. Simulated using COMSOL [21] for different gaps between the metal wires and the resonator edge, which are controlled by depositing thicker sacrificial PECVD $SiO_2$ layer. (Right) Schematic of the cross section of the tunable suspended micro-resonator describing how the suspended wire bridges cross over the resonator edge.

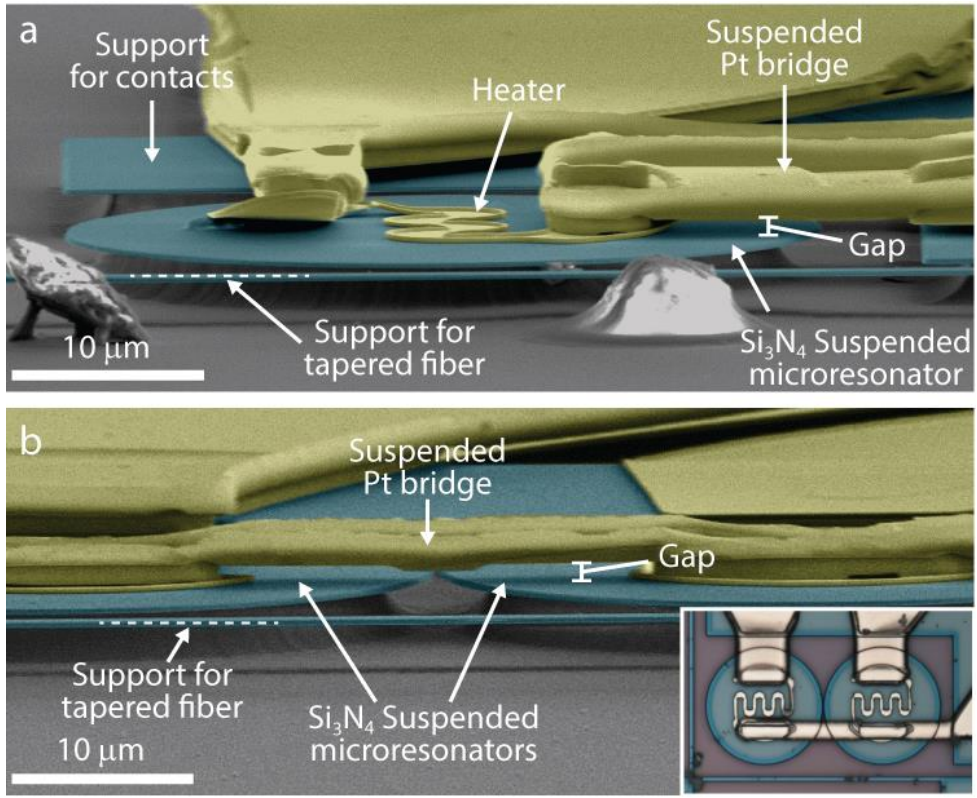

Fig. 4. False-colored scanning electron microscope image of the fabricated tunable suspended microresonators. (a) A single suspended microresonator (blue disk) with the proposed integrated heater and suspended Pt wire bridges (yellow) are shown. The Pt bridges are fully released, maintaining a small gap between the metal and the resonator edge for optical isolation. (b) A close-up of the suspended Pt wires in an evanescently coupled suspended microresonators array is shown. A common ground connects both heaters for the evanescently coupled suspended microresonators. Inset is an optical microscope image of device in (b).

## 3. Experimental Results

We demonstrate tuning of the resonance of a 40 μm-diameter suspended microresonator by 9.7 GHz using 5.3 mW of electrical power on the heaters without affecting the optical Q. Fig. 5 is the measured transmission spectra of the suspended microresonator at various heater powers. The initial resonant wavelength (heater power = 0 W) is 1551.13 nm. As the heater power increases, the resonance redshifts due to the thermo-optic effect. The temperatures of the resonator at various heater powers are extracted from the resonance detuning using the thermo-optic coefficient of $Si_3N_4$ ($dn_{Si_3N_4}/dT = (2.45\pm0.09) \times 10^{-5}$ RIU / °C) [23]. From the measured loaded Q at each resonator temperature in Fig. 5, we observe no degradation of the Q, confirming that the Q is not affected by heating. This implies that the wires do not buckle down and add loss to the resonator as heater power increases. The slight increase in the loaded Q and the change in transmission extinction observed at higher temperatures are likely due to slight drifts in the taper-resonator coupling conditions [24]. We measure the heater response to be in the order of 30 μs (see Fig. 9 in *Appendix*). Our tuning efficiency is 350 MHz / K, or 15 nm / W, which is more efficient by a factor of 15 than previously demonstrated $Si_3N_4$ tuning [12]. To achieve even higher tuning efficiency, heater design may be optimized to increase the heating area and breakdown power.

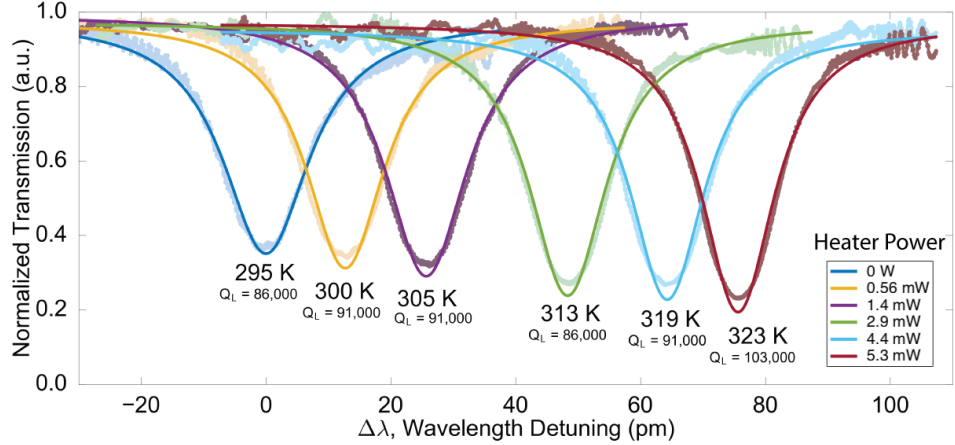

Fig. 5. Resonance detuning of the suspended microresonator with our integrated approach. The resonance wavelength at no heater power, initial detuning, is 1551.13 nm. For each spectrum, the solid line is the Lorentzian fit to the measurement. The temperatures of the resonator are extracted from the resonance detuning using the thermo-optic coefficient of $Si_3N_4$ [23]. The loaded Q factor is measured at each temperature, and does not decrease as we detune the resonance, demonstrating that the Q factor is not affected by the heating. The slight increase in the loaded Q and the change in transmission extinction observed at higher temperatures are likely due to slight drifts in the taper-resonator coupling conditions [24].

In order to demonstrate the scalability and the ability to control the coupling strength of multiple resonators, we apply our approach to an array of two evanescently coupled suspended optical resonators (refer to Fig. 1(c) for a schematic of this type of system). The evanescently coupled suspended resonator system is important in various applications [5–7], but the coupling strength could not be dynamically controlled. The resonators are difficult to couple optically as the optical resonant frequencies are prone to fabrication variations. Even variations on the nanometer scale – typically observed during fabrication – lead to different resonant frequencies (e.g., 5.4 GHz of resonance difference in Ref. [5]). Therefore, sets of evanescently coupled resonators are unlikely to exhibit maximum coupling strength because the resonances do not perfectly align.

We fabricate two identically designed and evanescently coupled suspended resonators, and tune the appropriate resonator to align the resonances and maximize optical coupling. Fig. 6 shows the spectra of the coupled suspended resonators (the left resonator, L, and the right resonator, R) at different detuning. The light is coupled to R with a fiber taper as shown in the bottom left infrared (IR) image in Fig. 6. The two resonators initially have different resonances due to fabrication variability; the resonances are centered around 1556.33 nm and 1556.34 nm for L and R, respectively, for heater power of 0.72 mW. As the resonance difference (1.2 GHz) is greater than the coupling rate between the two resonators, the light resonates in R much longer than in L, and R resonance has a much larger extinction than the L resonance. The resonance of L is blue-detuned relative to that of R, so we redshift the L resonator using its heater. As we increase the temperature of L, the L resonance redshifts. As the resonance difference becomes comparable to the coupling rate, the transmission extinction increases, indicating increased coupling strength due to better resonance alignment. The resonances best align at around heater power equal to 3.28 mW, which is centered around 1556.35 nm (dark green spectrum in Fig. 6). From the middle IR camera image in the left of Fig. 6, one can see that light is circulating within both resonators even if the fiber taper is coupling light into R. We further redshift the L resonance to cross over the R resonance and observe the expected anti-crossing behavior of a coupled resonator system with less than 7 mW of heater power. Our tuning mechanism, therefore, can overcome unavoidable fabrication variations and dynamically control the coupling strength between the two coupled resonators.

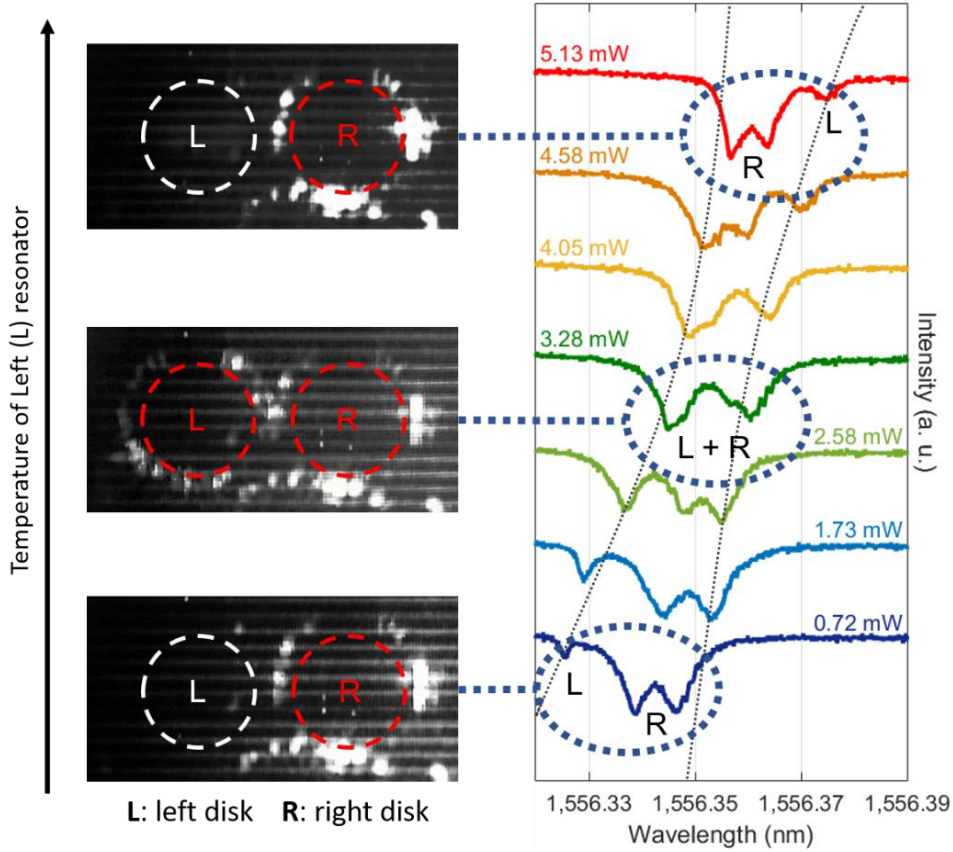

Fig. 6. Controlling the coupling strength between two suspended microresonators. (Left) Images of the resonators are taken with an IR camera to ensure optical coupling between the resonators (see inset of Fig. 4(b) for an optical microscope image of the devices). (Right) As fabricated, the resonances of the device are centered around 1556.33 nm and 1556.34 nm for the left resonator (L) and the right resonator (R), respectively (dark blue spectrum, heater power = 0.72 mW). The resonance of R is split in 0.72 mW spectrum due to the existence and coupling between clockwise and counter-clockwise propagating mode (see *Appendix*). We tune the L resonator using its heater to induce a red thermo-optic shift. As we increase the temperature of L, the resonances of both resonators align (dark green spectrum, heater power = 3.28 mW) and from the IR camera image one can see that both resonators are in resonance. We further increase the temperature of L so that the L resonance crosses over the R resonance. We observe this full anti-crossing behavior with less than 7 mW of heater power.

Our resonance tuning of the evanescently coupled resonator array, where we tune one of the resonators (L), agrees well with the coupled resonator model [25], indicating our ability to achieve maximum coupling efficiency between the resonators. The resonance detuning of evanescently coupled resonators is described by:

$$\omega_{s,as} = \omega_{avg} \pm \sqrt{\frac{\Delta\omega^2}{4} + \gamma^2}, \qquad (1)$$

where, $\omega_{avg} = (\omega_L + \omega_R)/2$, is the average resonance frequency of the two resonators, $\Delta\omega = \omega_L - \omega_R$, is the relative detuning between the two resonances, and $\gamma$ is the coupling rate between the two resonators. The L and R resonances redshift with increasing L heater power, $P$, according to $\omega_{L,R} = \omega_{0L,R} - \eta_{L,R} P$, where $\omega_{0L,R}$ is the initial resonant frequency of the L or R resonator and $\eta_{L,R}$ is the tuning efficiency of L or R resonator. Fig. 7 describes how the peak wavelengths of the resonators evolve as we increase the heater power of the L resonator. At different heater powers, we identify the peak positions of each resonator then sweep the laser wavelength across the R resonance to take the IR images of the inset. The IR images provide an independent confirmation of which resonator is in resonance at different coupling strengths. The red points are peaks of the two resonators extracted from the measured spectra in Fig. 6. The dotted black lines are guides to the eye that represent resonance detuning of L and R resonances when there is no coupling, $\gamma = 0$. We first fit the measured peaks to the dotted black lines to extract the tuning efficiency, $\eta_L$ and $\eta_R$, of the resonators, which are $2\pi \times 1.6$ MHz / W and $2\pi \times 28$ kHz / W, respectively. The R resonance detunes despite injecting power into the L heater, as observed by the small redshift of the R resonance, due to small thermal crosstalk between the two resonators through air thermal conduction and/or their common electrical ground. Similar thermal crosstalk between the two resonators is also observed when tuning cladded devices [11]. The tuning efficiency $\eta_R$ is included in the modeling to consider such thermal crosstalk. The blue lines are the resonance detuning of the symmetric and antisymmetric supermodes according to Eq. (1). From the fit, the coupling rate is extracted as $\gamma = 2\pi \times 960$ MHz.

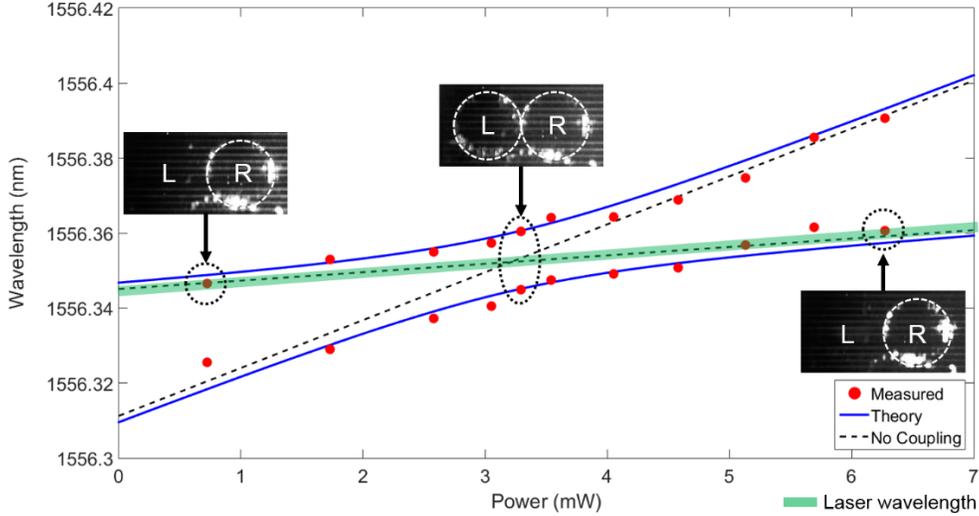

Fig. 7. Anti-crossing of the coupled resonators when redshifting the L resonance to cross the R resonance using less than 7 mW of heater power. The red points are peaks of the two resonators extracted from the measured spectra in Fig. 6. The blue line is the simulated lines according to Ref. [25] and Eq. (1), including the small thermal crosstalk between the two resonators. The dotted black lines are guides to the eye that represent resonance shifts of each resonator where there is no coupling. The laser wavelength (green line) is swept across the R resonant wavelength to take the inset IR images for independent confirmation of which resonator is in resonance.

## 4. Conclusions

We demonstrate the active tuning of traditionally passive suspended microresonators in single and evanescently coupled configurations utilizing suspended wire bridges and microheaters. Our approach enables scalability of suspended microresonators for complex on-chip photonics applications as we tune them in an integrated manner. The suspended wires connect the heaters to the contacts isolated from the resonators without requiring a cladding layer, enabling integrated control of the heaters without macroscopic probes directly interacting with the resonators.

## 5. Appendix

### 5.1 Resonance splitting in an array of evanescently coupled resonators

We simulate the evanescently coupled resonator system in order identify different types of resonance splitting in our fabricated coupled resonator system. Fig. 8 shows the configuration of the simulated coupled resonator system with various coupling parameters, $\beta$ and $\gamma$, as well as simulated transmission spectra. Light is coupled to the first resonator from a bus waveguide, or a fiber taper. Due to evanescent coupling, the two resonators form a new eigenmode system, in which the two eigenmodes, or supermodes, are a symmetric and anti-symmetric mode. These supermodes are hybridization of clockwise (CW) and counter-clockwise (CCW) propagating modes within each resonator. The hybridized supermodes span both resonators and the resonant frequencies are split as shown in Fig. 8(b) and 8(c). The splitting is larger if the coupling strength between the resonators, $\gamma$, is greater (i.e. the distance between the resonators is smaller). Within each supermode (symmetric or anti-symmetric), there is another resonance splitting.

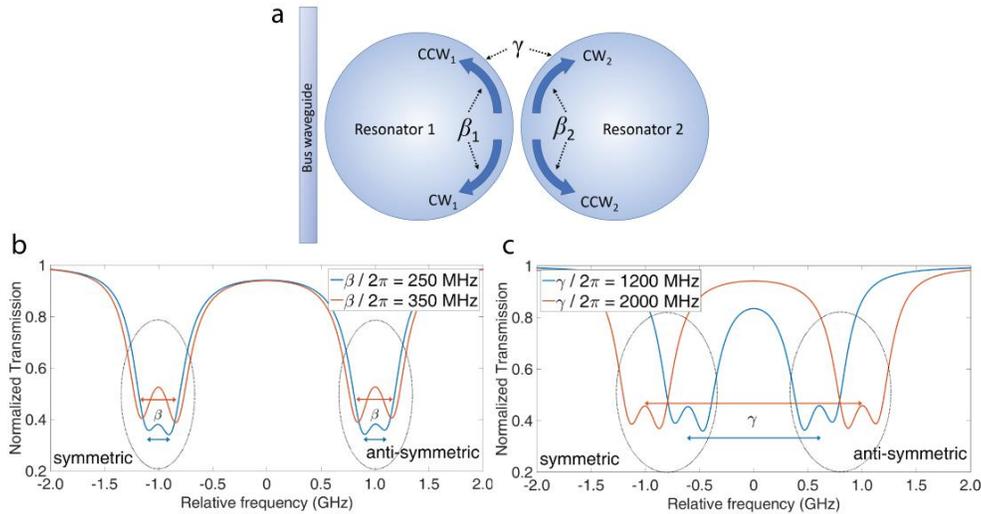

Fig. 8. Coupling dynamics in a system of evanescently coupled resonators. (a) There are different types of coupling. In each resonator, there is a coupling between clockwise (CW) mode and counter-clockwise (CCW) mode, $\beta_{i=1,2}$, due to side-wall scattering or other perturbation that splits the resonance of each resonator. In addition, the CW and CCW modes in the two resonators hybridize into symmetric and anti-symmetric supermodes spanning both resonators with coupling strength $\gamma$ due to evanescent coupling. (b) The resonance splitting caused by $\beta_i$, the coupling between CW and CCW modes, is shown at different coupling strengths with fixed evanescent coupling rate, $\gamma$. (c) The resonance splitting caused by $\gamma$, the coupling between symmetric and anti-symmetric supermodes, is shown at different coupling strengths with fixed CW and CCW coupling rate, $\beta_i$.

This is due to the coupling between CW and CCW mode of each resonator with coupling strength $\beta_i$ ($i$ = Left or Right resonator). This coupling between CW and CCW modes occurs through scattering from the imperfections of the resonator sidewall or the perturbation of the neighboring resonator. As shown in Fig. 8(a), the splitting within each supermode is larger if the coupling strength $\beta_i$ is greater. This coupling between CW and CCW modes is usually weak in an isolated single resonator, but in a high-Q resonator or in a system of evanescently coupled resonators, the side wall roughness or the neighboring resonator's presence introduces a strong perturbation such that CW and CCW modes coupling becomes stronger. The resonance splitting of the R resonance in the dark blue spectrum (0.72 mW) in Fig. 6 is due to the coupling between the CW and CCW mode. Also, as the two resonators (L and R) have optical resonant frequencies that is greater than their coupling strength, $\gamma$, there is a large difference in extinction between the two resonances. When the L resonance redshifts and the resonances of the two resonators are aligned, the resonance still splits. This is now primarily due to the coupling of supermodes with coupling strength $\gamma$ as described in Fig. 8(a). Because the coupling between CW and CCW modes ($\beta_i$) is usually weaker than the coupling between the supermodes ($\gamma$), the resonance splitting in the dark green spectrum (3.28 mW) of Fig. 6 is greater than that of R resonance in the dark blue spectrum (0.72 mW).

*5.2 Heater response time*

We measure the heater response time on the 40-μm resonator to be in the order of 30 μs. The heater is driven with a 5 kHz square wave with 1.25 V peak-to-peak amplitude. Fig. 9 shows the square wave driving the heater in blue line and the optical transmission response of the resonator at a fixed wavelength in orange line. The microheater is capable of modulation speed of 33 kHz, which is sufficient to stabilize cavity dynamics.

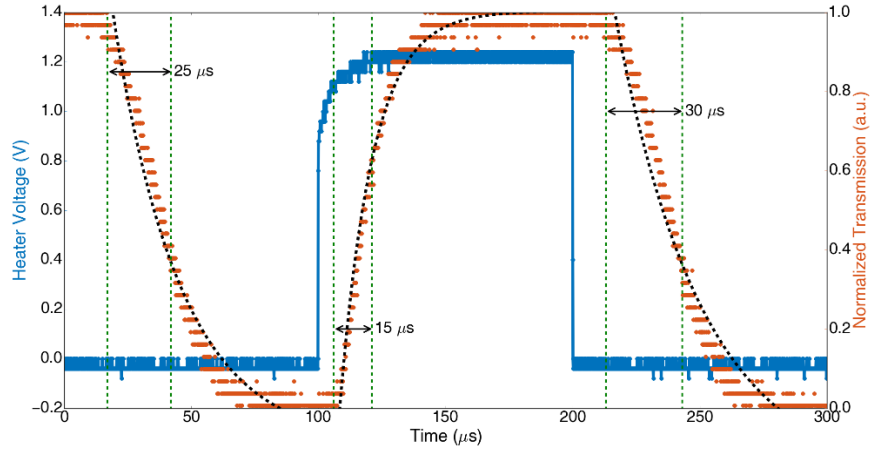

Fig. 9. Measured heater response time on the 40-μm resonator. The heater is driven with 5 kHz square wave with 1.25 V peak-to-peak. The left vertical axis is the heater voltage and the right vertical axis is the optical transmission of the resonator at a fixed wavelength. Rising and falling time are around 25 μs and 15 μs, respectively.

*5.3 Intrinsic quality factor, Q*

We measure the intrinsic Q of the L resonator in the evanescently coupled resonator array to confirm that the microheater and suspended metals do not induce loss in the resonator. Fig. 10 shows the transmission spectrum of the L resonator extremely undercoupled to the fiber taper. The intrinsic Q is around $5.2 \times 10^5$ as shown in Fig. 10. This intrinsic Q is comparable to other $Si_3N_4$ suspended microresonators fabricated with the same procedures but without our tuning circuit [5], confirming that the tuning circuit does not induce loss on the resonator. Therefore, our approach minimally perturbs the optical dynamics of the resonator.

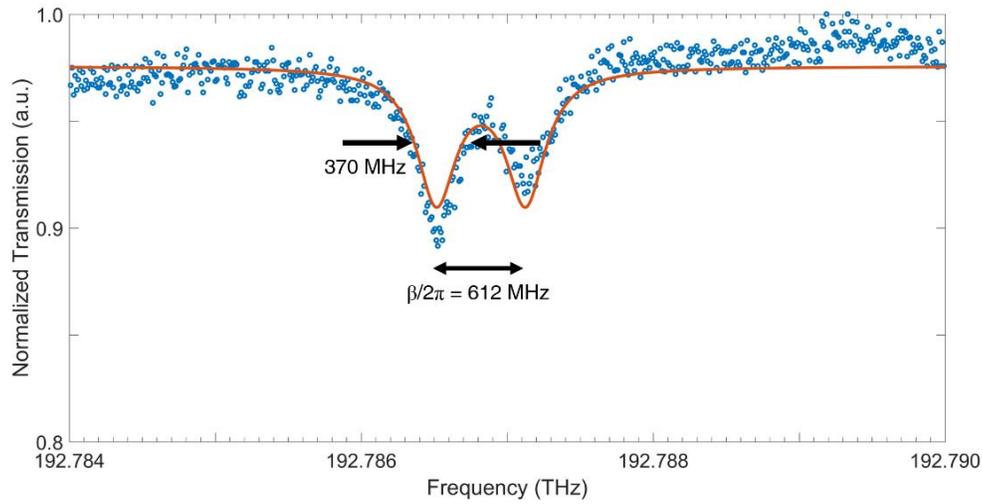

Fig. 10. Intrinsic Q measurement of the L resonator in the evanescently coupled system of the inset of Fig. 4(b). The intrinsic Q is around $5.2 \times 10^5$, comparable to $Si_3N_4$ suspended microresonators fabricated without the heating circuit [5]. This implies that the heaters and the suspended wire bridges do not introduce optical loss. The coupling strength ($\beta$) between CW and CCW mode is about $2\pi \times 610$ MHz.

**Funding**

This work was also supported by the NSF through CIAN ERC under grant EEC-0812072.

**Acknowledgments**

The authors acknowledge Dr. Shreyas Shah and Dr. Jaime Cardenas for valuable fabrication help. This work was performed in part at the Cornell NanoScale Facility, a member of the National Nanotechnology Infrastructure Network, which is supported by the National Science Foundation (NSF) through grant ECCS-0335765, and made use of the Cornell Center for Materials Research Shared Facilities which are supported through the NSF MRSEC program (DMR-1120296). The SEM images were taken at the City University of New York Advanced Science Research Center NanoFabrication Facility.